\documentclass[conference]{IEEEtran}

\usepackage{amsmath}
\usepackage{tabularx}
\usepackage{booktabs}

\usepackage{graphicx}

\IEEEoverridecommandlockouts

\ifCLASSINFOpdf
\else
\fi
\begin{document}
\title{Compiler Phase Ordering as an Orthogonal Approach for Reducing Energy Consumption \thanks{Parts of this paper were originally presented at the 19th Workshop on Compilers for Parallel Computing (CPC 2016), July 6-8 2016, Valladolid, Spain.}}

\author{\IEEEauthorblockN{Ricardo Nobre}
\IEEEauthorblockA{Faculty of Engineering,\\
University of Porto, and\\
INESC TEC, Porto, Portugal\\
Email: ricardo.nobre@fe.up.pt}
\and
\IEEEauthorblockN{Lu\'{i}s Reis}
\IEEEauthorblockA{Faculty of Engineering,\\University of Porto, and\\
INESC TEC, Porto, Portugal\\
Email: luis.cubal@fe.up.pt}
\and
\IEEEauthorblockN{Jo\~{a}o M. P. Cardoso}
\IEEEauthorblockA{Faculty of Engineering,\\
University of Porto, and\\
INESC TEC, Porto, Portugal\\
Email: jmpc@acm.org}}

\maketitle

\begin{abstract}
Compiler writers typically focus primarily on the performance of the generated program binaries when selecting the passes and the order in which they are applied in the standard optimization levels, such as GCC --O3. In some domains, such as embedded systems and High-Performance Computing (HPC), it might be sometimes acceptable to slowdown computations if the  energy consumed can be significantly decreased. Embedded systems often rely on a battery and besides energy also have power dissipation limitations, while HPC centers have a growing concern with electricity and cooling costs. Relying on power policies to apply frequency/voltage scaling and/or change the CPU to idle states (e.g., alternate between power levels in bursts) as the main method to reduce energy leaves potential for improvement using other orthogonal approaches.
In this work we evaluate the impact of compiler pass sequences specialization (also known as compiler phase ordering) as a means to reduce the energy consumed by a set of programs/functions when comparing with the use of the standard compiler phase orders provided by, e.g., --OX flags. We use our phase selection and ordering framework to explore the design space in the context of a Clang+LLVM compiler targeting a multicore ARM processor in an ODROID board and a dual x86 desktop representative of a node in a Supercomputing center.
Our experiments with a set of representative kernels show that there we can reduce energy consumption by up to 24\% and that some of these improvements can only be partially explained by improvements to execution time. The experiments show cases where applications that run faster consume more energy.
Additionally, we make an effort to characterize the compiler sequence exploration space in terms of their impact on performance and energy.
\end{abstract}

\IEEEpeerreviewmaketitle

\section{Introduction}

Mapping applications efficiently is very important when targeting systems with strict requirements (e.g., embedded systems, HPC), such as energy/power, performance, memory and/or storage.
Software optimization driven by an optimizing compiler helps to comply with requirements while using less resources in the process, contributing to the reduction of hardware costs and/or improving user experience.

An optimizing compiler sequence is a set of analysis or transformation passes that if orderly executed by a compiler (e.g. GCC \cite{gcc}, LLVM \cite{llvm}) over a tool specific representation of a function and/or program, will result in the generation of a representation of the same function/program optimized regarding a given metric.
The resultant representation typically assumes the form of machine code (e.g., object code or executable) for software compilation (which is the focus of the experiments presented in this paper); or an application-specific architecture described in a hardware description language (e.g., Verilog, VHDL) in the case of hardware compilation.

Compilers are typically distributed with a set of standard compiler optimization levels represented by flags.
These flags represent fixed compiler sequences and are typically tuned only for performance or code size.
Those sequences are accessible using flags such as --O1/--O2/--O3 for performance, and --Os for code size.
The existence of such flags is undeniably useful, as their use results in compiled code that not only tends to runs faster, but also tends to be more energy efficient.
Programmers typically rely on the standard compiler optimization levels to optimize their functions/applications.

Optimizing for performance and optimizing for energy efficiency are closely related.
In a situation where power is kept constant then a faster program is always a more energy efficient program.
However, power is not constant.
Nowadays, processors use a wide range of mechanisms in order to make computations more efficient. Power depends on frequency/voltage scaling (also known as CPU throttling), a technique for conserving power. In microprocessors typically there exist frequency/voltage pairs (also known as P-states).

Power depends on what operating/idle states (C-states, from C0 to C6 on Intel CPUs) are active at any given time.
The processor alternates between an operating state (e.g., C0) and idle states (e.g., C1 to C6), that result in the internal and/or external clocks being halted, and/or the voltage being further reduced and/or the turning off the cache memory.
Additionally, the activity in different processor components, such as the SIMD units (e.g., AVX2), memory caches, cores, can result in increased power consumption.
Finally, other components such as the system RAM also have variable energy consumption patterns depending on their usage, which depends on a number of factors. For instance, CPU cache misses increase the energy consumption.

Because of these aspects, optimizing for performance may not be fully in line with optimizing for energy; specially when relying on fixed compiler sequences such as the ones represented by the --OX flags.
Even if it was the case that energy efficiency always improved in direct relation with performance (i.e., 2$\times$ faster would equal 2$\times$ less energy), the use of the --OX flags (which are typically tuned for performance) would still leave potential for better performance and energy efficiency through the use of compiler sequences specially tailored for the given function and target platform pair.
No fixed compiler sequence can result in the best possible output code for all input functions/applications, even if considering a single target.

Phase selection and/or phase orders specialized for the input source code, non-functional requirements, and target, can lead to better software implementations. Phase selection deals with the selection of which compiler passes are executed in a given fixed compiler pipeline. A phase order is a set of analysis/optimization/lowering compiler passes executed in a given order.

In this paper we experimentally show that optimizing for performance through compiler-driven software optimization as a means of optimizing for energy does not always lead to the most energy efficient compiled functions/programs.
We achieve this goal by compiling multiprocessor-ready versions of 12 PolyBench/C \cite{polybench} kernels, to a dual Intel Xeon workstation, representative of a supercomputer node and to an ARM-based ODROID XU+E board, representative of hardware present on a mobile phone or tablet,
and comparing how execution time and energy consumption are affected for the execution of the binaries generated from compilation with phase orders generated by a design space exploration (DSE) method.
The results show that although improving performance tends to improve energy efficiency, there are a number of situations where the compiler sequences that result in the best performance do not translate into the best energy consumption.
In such cases, there are compiler sequences that allow achieving even better energy savings.

The rest of the paper is organized as follows.
Section \ref{section-experiments} explains the methodology of the experiments presented in this paper.
Section \ref{section-results} presents the experimental results.
Related work is presented in Section \ref{section-relatedwork}.
Final remarks about the presented work and ongoing work are presented in Section \ref{section-conclusions}.

\section{Experiments}\label{section-experiments}

We performed a number of experiments in two relevant target platforms in order to evaluate the impact of compiler optimizations using specialized compiler phase orders on both energy and performance.

\subsection{Platforms}
We consider two systems.
A workstation with two Intel Xeon E5-2630V3 CPUs (@2.4 - 3.2 GHz Turbo), and 128 GB of DDR4 (@2133 MHz), representative of a supercomputer node; and an ODROID XU+E single board computer~\cite{odroid}, with a Samsung Exynos~5410 SoC (part of the Exynos~5 Octa series), the same SoC in the the Samsung Galaxy S4 smartphone.
The Xeon E5-2630V3 is an X86-64 CPU with Intel's latest microarchitecture for the workstation/server market, the Haswell-EP microarchitecture.
The Exynos 5410 SoC on the ODROID includes a Cortex-A15 1.6 GHz quad core CPU and a 1.2 GHz Cortex-A7 quad core CPU, in a configuration referred to as \emph{big.LITTLE}, and 2 GB of LPDDR3 on the same package.
The \emph{big} cores are designed for maximum compute performance, while the \emph{LITTLE} cores are designed for maximum power efficiency.
Unlike a traditional 8-core CPU (or a dual 4 core), the \emph{big.LITTLE} configuration means that the \emph{big} (Cortex-A15) and the \emph{LITTLE} (Cortex-A7) CPU cores take turns to execute a task, which is migrated between the two types of cores during execution, in a joint effort to make computation more power and energy efficient.

The operating system for both platforms is Ubuntu.
On the dual Xeon-based workstation we use a 64-bit Ubuntu 16.04 LTS system with Linux kernel 4.4.0, and on the ODROID board we use Ubuntu 14.04.2 LTS with Linux kernel 3.4.75.

For experiments with OpenMP, we use version 3.7.1 of the LLVM OpenMP runtime \cite{llvmopenmp} in both platforms.

For the dual Xeon E5 V3 platform we consider the default power settings with Turbo-boost activated to effectively drive some of the CPU cores up to 3.2 GHz from the base clock of 2.4 GHz.
For the ODROID XU+E board we use the two types of ARM cores (Cortex-A15 @1.2GHz and Cortex-A7 @1.6 GHz) in the \emph{big.LITTLE} configuration.

In both cases, frequency voltage scaling is activated and managed by the default power governor on the Linux distributions used; the \emph{powersave} power governor for the Dual Xeon and the \emph{on-demand} power governor for the ODROID.

\subsection{Functions}
In this experiments we use 12 kernels from PolyBench \cite{polybench} (version 4.1), and generated parallel versions of the kernels with PLUTO \cite{pluto}, a tool for automatic parallelization.
PLUTO relies on a polyhedral model generated from the input function/program, specially the parts concerned with loops and operations with arrays, as an abstraction to safely perform parallelization of loops thorough annotating the code with OpenMP pragmas (coarse-grained parallelism) and apply other high-level transformations, such as tiling loops to improve locality (reduces cache misses), and vectorization (i.e., use of SIMD units such as AVX).

Table \ref{table-functions} depicts the functions (and input data) used for the experiments and their number of lines of code (excluding comments) for both the original C version and the OpenMP annotated versions generated using PLUTO.
Parallel version of functions identified by an asterisk (*) were generated with loop tiling (\emph{--tile} option) in addition to parallelization (using the \emph{--parallel} option), which annotates the code with OpenMP pragmas.

\begin{table}[tbp]
\centering
\caption{Description of 12 \mbox{PolyBench/C} 4.1 functions, input parameters, and lines of code for the original implementations and OpenMP versions generated with the PLUTO automatic parallelizer.}
\label{table-functions}
\setlength\tabcolsep{3pt} 
\begin{tabularx}{\columnwidth}{@{}l p{3.25cm} X c@{}}	
\toprule
\parbox{1cm}{\centering \textbf{Function}}       & \parbox{3cm}{\centering \textbf{Description}}   & \parbox{1.5cm}{\centering \textbf{Input}}  & \parbox{1cm}{\centering \textbf{CLOC}}                                                                                                   \\ \midrule
2mm		& Autocorrelation of an input vector.    	& \emph{ni: 400, nj:450, nk:550, nl: 600} & 11 / (156)                                                                         \\ \hline
3mm		& Endian-swap a block of 16-bit values.		& \emph{ni: 400, nj:450, nk:500, nl: 550, nm: 600} & 27 / (136)                                                                            \\ \hline
atax	& Endian-swap a block of 32-bit values. 	& \emph{n: 10800, \mbox{m: 10800}} & 31 / (73)                                                                         \\ \hline
correlation\textsuperscript{*}	& Endian-swap a block of 64-bit values.	& \emph{n: 1000, m: 800} & 39 / (174)                                                                            \\ \hline
doitgen\textsuperscript{*}      & Move block of memory.           		& \emph{nr: 120, n: 110, np:130}       & 13 / (62)                                           \\ \hline
gemver\textsuperscript{*}       & Vector product of two input arrays.	& \emph{n: 10000} & 7 / (67)                                                                             \\ \hline
jacobi-2d\textsuperscript{*}        & Dot product of two arrays.      & \emph{n: 650, \mbox{tsteps: 250}}       & 17 / (175)                                                                         \\ \hline
mvt        & Complex FIR.            & \emph{n: 6000}               & 24 / (29)                                                                         \\ \hline
nussinov \textsuperscript{*}          & Least Mean Square Adaptive Filter.  & \emph{n: 1100}   & 17 / (57)                                                                        \\ \hline
seidel-2d           & Convert IEEE FP into Q.15 format.  & \emph{n: 800, \mbox{tsteps: 200}}   & 16 / (27)                                                     \\ \hline
syr2k          & Matrix Multiply.           & \emph{n: 600, m:500}            & 19 / (44)                                                                        \\ \hline
syrk        & Transposes a matrix of 16-bit values. & \emph{n: 800, m:700} & 8 / (43)          \\                           
\bottomrule
\end{tabularx}
\end{table}

\subsection{Energy and Performance Measurements}

For the experiments presented in this paper we measured the energy consumed by CPU and RAM on each platform. The energy values reported are always for the sum of both.

For measuring energy on the Intel system we use the Running Average Power Limit (RAPL) interface, which provides access to a mechanism to regulate power usage and to a set of registers with power and energy measurements.
Power measurements are generated on the fly, per socket, using an on-chip energy model that relies on hardware performance counters and I/O.
RAPL is available in Intel CPUs starting with the Sandy Bridge microarchitecture, and has been shown to correlate well with real measurements or at least to be indicative of overall energy and power trends \cite{sandy, haswell}.
RAPL gives access to four domains, \emph{Package}, which includes CPU cores, memory cache, memory controller, \emph{PPO} (CPU cores), \emph{PP1} (GPU) and \emph{DRAM}.
For our experiments we evaluate energy consumption as the sum of the energy consumption reported for the \emph{Package} and the \emph{DRAM} domains of each CPU.

Although RAPL has been found to underestimate DRAM energy use on some architectures, this does not impact Haswell-EP CPUs, as this architecture uses a different estimation mechanism that includes actual measurements~\cite{haswell}.
Haswell-EP does not support PP0~\cite{haswell}, but that makes no difference for these experiments because we want to take into account the energy used by cache and the memory controller in addition to the energy used by the CPU cores.
RAPL readings have a sampling frequency of close to one millisecond (1 KHz).

The Odroid XU+E single board computer features 4 energy sensors: ARM, KFC, MEM and G3D.
We query these sensors for energy measurements (except the G3D sensor), which refer to the A15 (big CPU), A7 (LITTLE CPU), memory and GPU subsystems, respectively.
For each of these, voltage, amperage and wattage are reported.
The update period of the energy sensors of the Odroid XU+E devices is 263,808 microseconds.	

We access the Intel RAPL measurements through the Linux \emph{perf\_event} subsystem, and the wattage measurements for the ODROID by querying the \verb|/dev/sensor_*| device files, using the \verb|ioctl| function.

For performance measurements we rely on \emph{clock\_gettime()} Linux calls, which compared with other calls for timing measurement (e.g., \emph{gettimeofday()}) allows higher precision and the ability to request specific clocks.
We use the the \emph{CLOCK\_MONOTONIC} clock.
The Linux function \verb|clock_getres()| reports one nanosecond of resolution in both platforms for the clock used.

\subsection{Datasets}
PolyBench provides \emph{mini}, \emph{small}, \emph{medium}, \emph{large}, and \emph{extra large} datasets. We selected one of these datasets on a function-by-function basis. 

For each function, we selected/costumized a dataset to make the execution time of each function (compiled with --O3) larger than 1 second on the ODROID XU+E board in order to have enough samples to obtain sufficiently precise energy measurements.

We use the same datasets (i.e., the ones we selected on the ODROID platform) with the Intel Xeon platform, resulting in execution times much smaller than on the ARM-based board, but still well above what is required for precise energy measurements with RAPL.

\subsection{Compilation and Validation}

\begin{figure} [tbp]
 \begin{center}
  \includegraphics[width=3.2in]{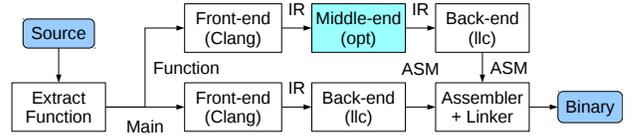}
 \end{center}
 \caption{Compilation flow for each source code.}
 \label{fig-flow}
\end{figure}

There are two compilation flows: one for the function to be optimized and other for the \emph{main()} function (and any other functions/code if existent).
We use Clang \cite{clang} to generate LLVM IR from the C code of the functions considered for optimization.
This LLVM IR is then passed to the LLVM Optimizer tool (the \emph{opt} tool) for optimization.
The transformed LLVM IR, after optimization with \emph{opt} using a compiler phase order passed as parameter (or one of the --OX flags), is then passed to the LLVM static compiler (the \emph{llc} tool) for generation of the assembly code for the target architecture (i.e., x86-64 for the Xeon CPUs and ARMv7 for the ARM CPUs).
The IR for the \emph{main()} function and any other functions, other than the function to optimize, are do not transformed by the \emph{opt} tool.
Figure \ref{fig-flow} shows the compilation flow.

As we compiled and executed the application on the same platform, we used the \emph{-mcpu=native} flag with the \emph{llc} tool.
This resulted in the generation of assembly code with vector instructions, NEON instructions for the ARM CPUs and AVX instructions for the Xeon CPUs. 
In addition, we used \emph{-fp-contract=fast} with \emph{llc} so that fused multiply-add machine instructions (supported by both the Intel and the ARM CPUs used) are used when possible.

In order to deal with a situation where the compiler tools get stuck (i.e., gets trapped in an infinite loop, for some combination of function, phase order and/or target), our DSE framework allows setting a time limit for each call to the tools (in this case, \emph{clang}, \emph{opt}, and \emph{llc}).
After some experiments with different values, we set this time limit to 10 seconds for the Xeon platform and 1 minute for the ODROID, which gives more than enough time for the tools to work with the input functions (or an LLVM IR representing them).
In the experiments with both targets, each binary generated by compilation is given 1 minute to execute.

The generated binaries are validated by comparing the output of a non-optimized serial version (i.e, as originally from PolyBench/C 4.1) of the function that is being optimized with the output of the optimized function. We classify as correct each output value if the absolute value of the difference between the two outputs (a positive floating point number) does not differ more than $0.001$.

For the validation step we always use the smallest available PolyBench/C dataset (i.e., the \emph{mini} dataset), in order to reduce the execution time needed for the validation.

\subsection{Compiler Phase Orders Exploration} \label{subsec:experiments-exploration}

We created a set of sequences by randomly generating 1,000 compiler sequences composed of 128 compiler passes each using our compiler phase selection/ordering exploration framework, using an ARC4-based pseudo-random number generator from \cite{seedrandom}.
For each position of a compiler sequence, we randomly selected a pass from the set of LLVM 3.7.1 passes presented in Table \ref{table-passes}.

\begin{table}[tbp]
\centering
\caption{LLVM Optimizer compiler passes used for exploration.}
\label{table-passes}
\setlength\tabcolsep{3pt} 
\begin{tabularx}{\columnwidth}{@{}c c c c@{}}
\toprule
-aa-eval            & -adce              & -add-dis.          & -alig.-f.-ass.       \\ \hline
-alloca-hoisting    & -always-inline     & -argpromotion      & -ass.-cache-track.   \\ \hline
-atomic-expand      & -barrier           & -basicaa           & -basiccg             \\ \hline
-bb-vectorize       & -bdce              & -block-freq        & -bounds-checking     \\ \hline
-branch-prob        & -break-crit-edg.   & -cfl-aa            & -codegenprepare      \\ \hline
-consthoist         & -constmerge        & -constprop         & -correlated-prop.    \\ \hline
-cost-model         & -count-aa          & -da                & -dce                 \\ \hline
-deadargelim        & -debug-aa          & -delinearize       & -die                 \\ \hline
-divergence         & -domfrontier       & -domtree           & -dse                 \\ \hline
-early-cse          & -elim-avail-ext.   & -extract-blocks    & -flattencfg          \\ \hline
-float2int          & -functionattrs     & -globaldce         & -globalopt           \\ \hline
-globalsmodref-aa   & -gvn               & -indvars           & -inline              \\ \hline
-inline-cost        & -instcombine       & -instcount         & -instnamer           \\ \hline
-instrprof          & -instsimplify      & -intervals         & -ipconstprop         \\ \hline
-ipsccp             & -irce              & -iv-users          & -jump-threading      \\ \hline
-lazy-value-info    & -lcssa             & -libcall-aa        & -licm                \\ \hline
-lint               & -load-combine      & -loop-accesses     & -loop-deletion       \\ \hline
-loop-distribute    & -loop-extract      & -loop-ex.-single   & -loop-idiom          \\ \hline
-loop-instsimpl.    & -loop-interchan.   & -loop-reduce       & -loop-reroll         \\ \hline
-loop-rotate        & -loop-simplify     & -loop-unroll       & -loop-unswitch       \\ \hline
-loop-vectorize     & -loops             & -lower-expect      & -loweratomic         \\ \hline
-lowerbitsets       & -lowerinvoke       & -lowerswitch       & -mem2reg             \\ \hline
-memcpyopt          & -memdep            & -mergefunc         & -mergereturn         \\ \hline
-mldst-motion       & -mod.-debuginfo    & -nary-reass.       & -no-aa               \\ \hline
-objc-arc           & -objc-arc-aa       & -objc-arc-apelim   & -objc-arc-contrac.   \\ \hline
-objc-arc-expand    & -pa-eval           & -part.-inliner     & -part.-inl.-libcal.  \\ \hline
-pl.-ba.-safe.-im.  & -place-safep.      & -postdomtree       & -prune-eh            \\ \hline
-reassociate        & -reg2mem           & -regions           & -rewr.-sta.-for-gc   \\ \hline
-rewrite-symbols    & -safe-stack        & -sancov            & -scalar-evolution    \\ \hline
-scalarizer         & -scalarrepl        & -scalarrepl-ssa    & -sccp                \\ \hline
-scev-aa            & -scoped-noalias    & -s.-c.-o.-f.-gep   & -simplifycfg         \\ \hline
-sink               & -slp-vectorizer    & -slsr              & -spec.-execution     \\ \hline
-sroa               & -strip             & -str.-dead-d.-info & -str.-d.-proto.      \\ \hline
-strip-d.-declare   & -strip-nondebug    & -structurizecfg    & -tailcallelim        \\ \hline
-targetlibinfo      & -tbaa              & -tti               & -verify      \\
\bottomrule
\end{tabularx}
\end{table}

This set of sequences was iteratively used to compile each of of the functions considered, resulting in the generation of up to $2,000$ binaries for each function considered in the experiments.
$1,000$ binaries for the OpenMP versions and $1,000$ binaries for the versions without OpenMP.
We note that some sequences failed to produce valid binaries when compiling some of the functions.
Energy consumption and execution time metrics were extracted from a single execution of each of those binaries, relying on the \emph{OMP\_NUM\_THREADS} environment variable to set the number of OpenMP threads for the OpenMP versions.

Then, for any given function, we built a model as in \cite{nobre2015, nobre2016} with the phase orders (generated in the previous step) for all other 11 PolyBench/C functions (one phase order per function) for the configuration (i.e., no OpenMP, or OpenMP with specific number of threads) that resulted in the lowest energy consumption when compiling with the standard optimization levels.
This model represents a sequence space from which not only the sequences used in its construction can be regenerated, but others sharing similarities with them can be generated as explained in \cite{nobre2016}.
We used the leave-one-out approach to validate, implying the construction of 12 different models per target platform by considering the best phase orders for each of the functions except the one being optimized at any given time.

We relied each of the models to iteratively generate 1,000 sequences (sequences generated between different models differ), and used them in the compilation of the function correspondent to the use of each specific model. 

Energy consumption and performance metrics were extracted from evaluating the binaries a single time, and we selected the 5\% (i.e, 50) compiler phase orders that resulted in the lowest energy consumption.
Then, we executed the binaries generated with those phase orders for 25$\times$ (the same number of times as for the --Ox experiments), and analyzed the energy consumption and performance of the binaries. 

We identified, for each function and target platform (i.e., dual Xeon and ODROID), the --OX flags that leads to the lowest energy consumption for each number of threads on the OpenMP versions and for the compilation without OpenMP.
Then we calculated, for each of the 50 compiler sequences previously selected, energy consumption and performance improvement ratios over those --OX configurations.
For instance, for the \emph{2mm} function on the dual Xeon workstation, the lowest energy consumption version is the one using OpenMP with 16 threads (as with most other functions on the Xeon target) and with the --O3 optimization level; therefore ratios presented in this paper.
For each function and target pair, we report in this paper the energy consumption and performance improvement ratios for the version using the same number of OpenMP threds (or without OpenMP).

\section{Results} \label{section-results}

We first present results of energy consumption, performance and average power for the functions compiled with the default optimization flags.
Including results for the kernels compiled without OpenMP support, and with OpenMP support considering different numbers of OpenMP threads.

In addition, we present results for exploration of compiler phase orders focusing on energy consumption, generated with the methodology explained in Section \ref{subsec:experiments-exploration}.
We analyze those results by comparing them with the energy consumption and performance of the binaries generated with the standard optimization level flags, and we also individually comment on the level of correlation between energy consumption and performance.

Energy and performance value pairs are obtained by averaging 25 executions of each of the binaries resulting from compilation of a given function with a specialized phase order (or a --Ox flag).

Table \ref{table-bestoptimandthreads} presents the compilation (i.e., which optimization flag) and execution configurations that lead to the lowest energy consumption, the best performance, or the lowest average power consumption.
Serial execution is represented by \emph{S} for the version compiled without OpenMP and \emph{1T} (i.e., OpenMP with a single thread) otherwise.

\begin{table}[tbp]
\centering
\caption{Best compiler flag and execution parameters for for each kernel and target platform.}
\label{table-bestoptimandthreads}
\setlength\tabcolsep{3pt} 
\begin{tabularx}{\columnwidth}{   p{2cm} |  X  X  X | X  X  X  }
\toprule
		& & \parbox{2cm}{\textbf{Dual Xeon}}  &  & & \parbox{2cm}{\textbf{ODROID}}  &  \\ \hline
	\textbf{Function} & \textbf{Energy} & \textbf{Perf.} & \textbf{Power} & \textbf{Energy} & \textbf{Perf.} & \textbf{Power} \\ \hline
	2mm & \textbf{16T -- O3} & 16T -- O1 & S -- O3 & \textbf{S -- O2} & 4T -- O1 & 1T -- O3 \\ \hline
	3mm & \textbf{16T -- O3} & 16T -- O3 & S -- O2 & \textbf{S -- O3} & 4T -- O2 & S -- O3 \\ \hline
	atax & \textbf{16T -- O3} & 32T -- O1 & 1T -- O3 & \textbf{S -- O3} & 4T -- O1 & S -- O3 \\ \hline
	correlation & \textbf{S -- O3} & 16T -- O1 & S -- O3 & \textbf{S -- O1} & S -- O1 & S -- O0 \\ \hline
	doitgen & \textbf{S -- O3} & S -- O3 & S -- O3 & \textbf{S -- O1} & S -- O1 & 1T -- O2 \\ \hline
	gemver & \textbf{32T -- O3 } & 32T -- O3 & 1T -- O3 & \textbf{S -- O3} & 4T -- O1 & S -- O1 \\ \hline
	jacobi-2d & \textbf{8T -- O1} & 8T -- O2 & S -- O1 & \textbf{S -- O2} & 4T -- O2 & S -- O0 \\ \hline
	mvt & \textbf{16T -- O2} & 16T -- O3 & 1T -- O1 & \textbf{S -- O2} & 4T -- O2 & 1T -- O1 \\ \hline
	nussinov & \textbf{16T -- O1} & 16T -- O1 & S -- O3 & \textbf{S -- O3} & 4T -- O1 & S -- O1 \\ \hline
	seidel-2d & \textbf{16T -- O3} & 32T -- O2 & S -- O1 & \textbf{2T -- O2} & 4T -- O0 & S -- O1 \\ \hline
	syr2k & \textbf{16T -- O2} & 16T -- O2 & S -- O2 & \textbf{S -- O2} & 4T -- O3 & S -- O2 \\ \hline
	syrk & \textbf{16T -- O1} & 32T -- O3 & S -- O1 & \textbf{S -- O3} & S -- O2 & S -- O1 \\
    \bottomrule
\end{tabularx}
\end{table}

\subsection{Energy and Performance with Standard Optimizations}

We present results for the use of 1, 2, 4, 8, 16 and 32 OpenMP threads on the dual Xeon and 1, 2 and 4 threads on the ODROID when executing the functions compiled with the standard optimization flags, as well as results for experiments without OpenMP.

\subsubsection{Dual Xeon}

\begin{figure*} [!t]
\setlength\abovecaptionskip{0pt}
 \begin{center}
  \includegraphics[width=7in]{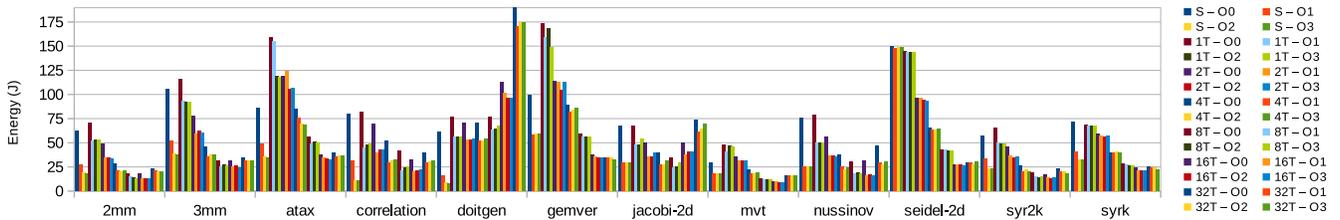}
 \end{center}
 \caption{Energy consumption in joules for each function when targeting the Dual Xeon with the standard optimization flags.}
 \label{fig-xeonenergy}
\end{figure*}

\begin{figure*} [!t]
\setlength\abovecaptionskip{0pt}
 \begin{center}
  \includegraphics[width=7in]{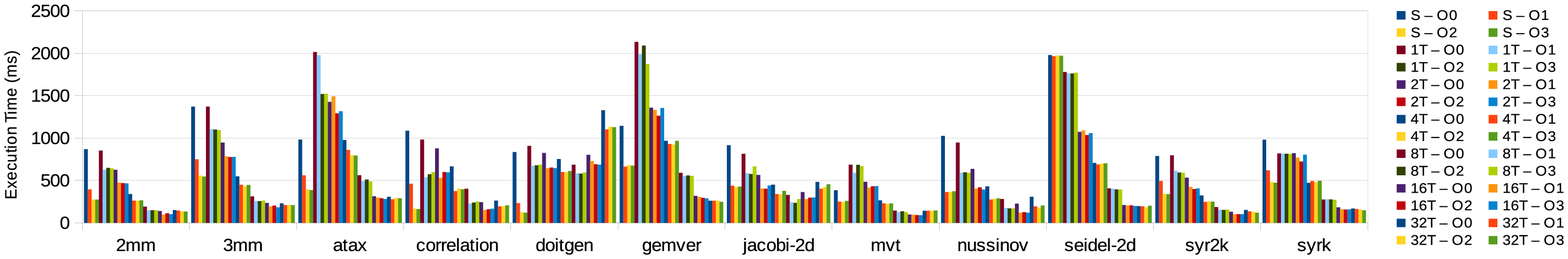}
 \end{center}
 \caption{Execution time in milliseconds for each function when targeting the Dual Xeon with the standard optimization flags.}
 \label{fig-xeonperformance}
\end{figure*}

\begin{figure*} [!t]
\setlength\abovecaptionskip{0pt}
 \begin{center}
  \includegraphics[width=7in]{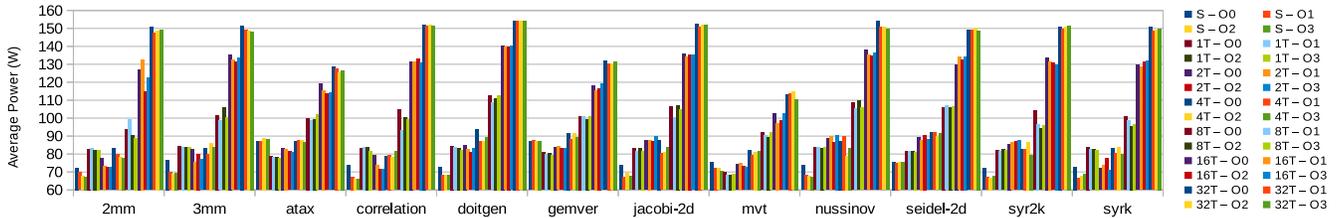}
 \end{center}
 \caption{Average power in watts for each function when targeting the Dual Xeon with the standard optimization flags. Calculated from energy consumption and execution time by $P = E / \Delta t$.}
 \label{fig-xeonpower}
\end{figure*}

\begin{figure*} [!t]
\setlength\abovecaptionskip{0pt}
 \begin{center}
  \includegraphics[width=7in]{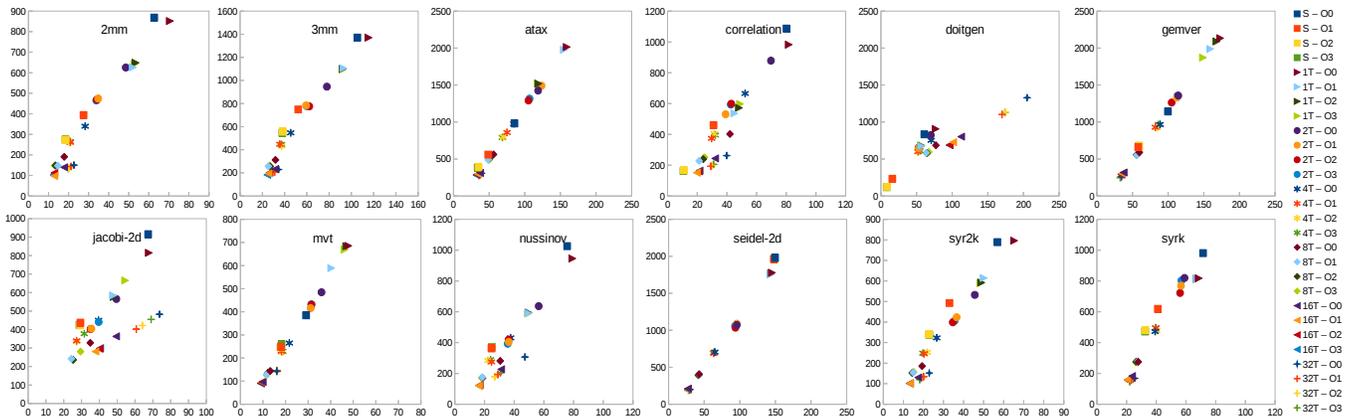}
 \end{center}
 \caption{Energy consumption in joules (horizontal axis) and execution time in milliseconds (vertical axis) on the Dual Xeon.} 
 \label{fig-xeonenergyperf}
\end{figure*}

Figures \ref{fig-xeonenergy}, \ref{fig-xeonperformance} and \ref{fig-xeonpower} depict the absolute energy consumption (in joules), the absolute performance (in milliseconds), and the average power consumption (in watts), on the dual socket Xeon platform for the execution with and without OpenMP of the 12 PolyBench/C functions considered for the experiments.
The functions were compiled with no optimization and with the -O1, -O2, -O3 standard optimization flags.
Figure \ref{fig-xeonenergyperf} represents energy consumption and execution time on the same chart for the binaries generated by compilation of the considered functions with the standard optimization levels.
Considering a number of execution threads, the use of the standard optimization levels (i.e., --O1, --O2, --O3) always results in the generation of binaries with both the lowest energy consumption and highest performance, when compared with the binaries generated without optimization.
The improvement obtained by using those flags is considerable, especially when OpenMP is not used (up to 87\% energy consumption reduction).
Additionally, when not using OpenMP --O2 and --O3 tend to improve energy consumption and performance over --O1.
The only exception are \emph{gemver} and \emph{seidel}.
The use of --O1 with \emph{gemver} results in saving 42\% energy (vs. not using any optimization level), but using --O2 or --O3 results in no additional improvement.
With \emph{seidel-2d} none of the optimization levels resulted in saving energy or improving performance.
When OpenMP is used the norm is that --O2 and --O3 do not tend to reduce the energy consumption or the performance further over --O1 by a significant amount, and in fact sometimes result in less energy efficient binaries. 
The binaries optimized with those optimization levels also tend to have lower average power consumption (\emph{atax}, \emph{gemver}, \emph{seidel-2d} are the exceptions) if OpenMP is not used.
The use of OpenMP resulted in performance degradation for some functions when using less than two threads.
Three of the most extreme cases were the \emph{atax}, \emph{correlation} and \emph{doitgen} functions.
The \emph{atax} and the \emph{correlation} functions need 16 threads for the OpenMP version to surpass the versions without OpenMP. 
Interestingly, the \emph{correlation} function without OpenMP (with --O2 or --O3) consumes $2\times$ less energy than any other version, making this a good example to show that energy and power are not always correlated.
The serial version without OpenMP of the \emph{doitgen} function is faster than any OpenMP version, independently on the number of threads used.
The use of 16 and especially 32 threads considerably negatively impacts energy consumption more than performance.
Other example where the use of more threads (16 or 32 threads) hurts energy consumption and performance disproportionally is \emph{jacobi-2d}.
Performance tends to improve over the serial non-OpenMP with the use of 4 or 8 OpenMP threads.
As expected, the average power tends to increase with the number of OpenMP threads.
The inverse behavior is seen for some functions when going from 1 to 2 OpenMP threads (\emph{2mm}, \emph{3mm}, \emph{correlation}, \emph{syrk}) or from 2 to 4 threads (\emph{jacobi-2d}, \emph{nussinov}, \emph{syr2k})
The serial versions without OpenMP tend to use less power, being \emph{atax}, \emph{gemver}, \emph{mvt} the exceptions.

\subsubsection{ODROID XU+E}

\begin{figure*} [!t]
 \begin{center}
  \includegraphics[width=7in]{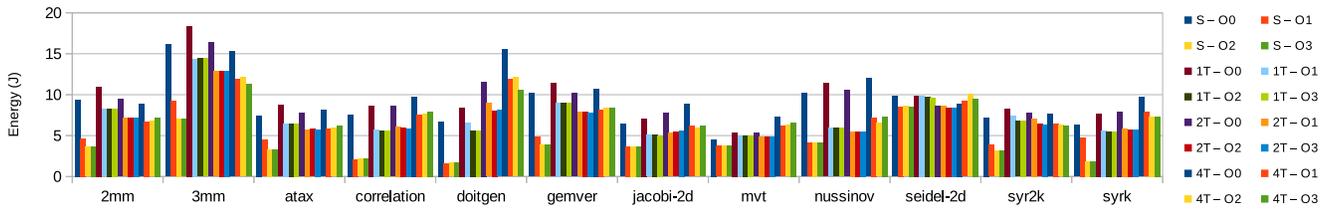}
 \end{center}
 \caption{Energy consumption in joules for each function when targeting the ODROID with the standard optimization flags.}
 \label{fig-odroidenergy}
\end{figure*}

\begin{figure*} [!t]
 \begin{center}
  \includegraphics[width=7in]{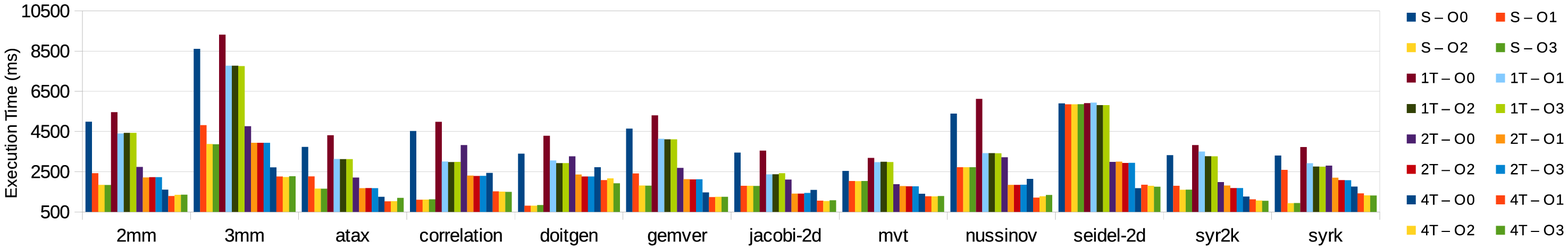}
 \end{center}
 \caption{Execution time in milliseconds for each function when targeting the ODROID with the standard optimization flags.}
 \label{fig-odroidperformance}
\end{figure*}

\begin{figure*} [!t]
 \begin{center}
  \includegraphics[width=7in]{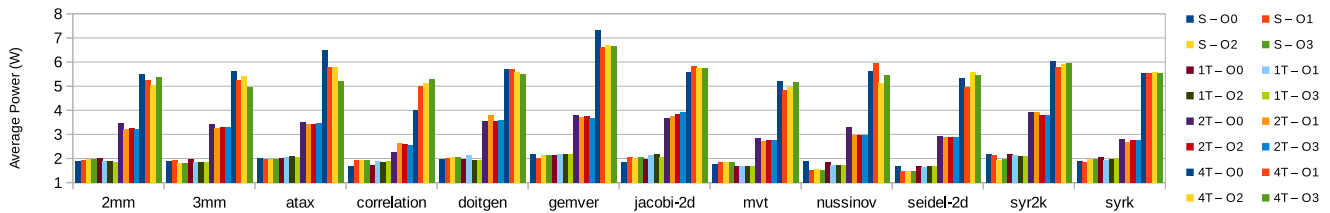}
 \end{center}
 \caption{Average power in watts for each function when targeting the ODROID with the standard optimization flags.}
 \label{fig-odroidpower}
\end{figure*}

\begin{figure*} [!t]
 \begin{center}
  \includegraphics[width=7in]{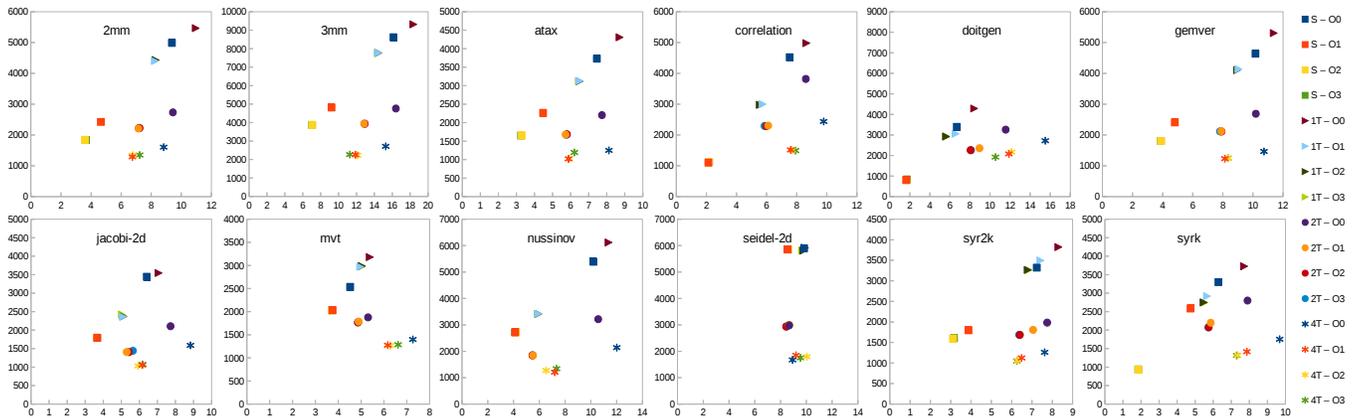}
 \end{center}
 \caption{Energy consumption in joules (horizontal axis) vs. execution time in milliseconds (vertical axis) on the ODROID.}
 \label{fig-odroidenergyperf}
\end{figure*}

Figures \ref{fig-odroidenergy}, \ref{fig-odroidperformance} and \ref{fig-odroidpower} depict energy consumption, performance and power on the ODROID.
Figure \ref{fig-odroidenergyperf} represents energy consumption and execution time on the same chart for the binaries generated by compilation of the considered functions with the standard optimization levels.
As with the Xeon-based platform, for the same function configuration (i.e., without OpenMP or with a given number of OpenMP threads), on the ODROID the use of --O1, --O2 or --O3 tends to result in lower energy consumption (the non-optimized versions never consumed less energy).
The only exception is \emph{seidel-2d} with 4 OpenMP threads, in which case energy consumption increases with the use of any optimization level.
Performance and energy consumption changes with use of optimization levels are closely related. If energy consumption decreases, then wall time tends to decrease.
This these changes do not always happen in the same proportion for both metrics.
For instance, \emph{seidel-2d} without OpenMP consumes up to 13\% less energy if compiled with the standard optimization levels, while improving less than 1\% in terms of performance in comparison with not relying on any of the optimization levels.
Compilation without OpenMP resulted in the generation of binaries that use less energy, with the \emph{seidel-2d} function being the only exception.
For this function the binary compiled with OpenMP saves more energy if executed with two threads.
The ODROID consumes less energy than the dual Xeon workstation and
the dual Xeon based system executes the compiled functions much faster than the ARM processors on the ODROID. 
The average power consumption on the ODROID is also much lower, with all functions executing bellow 8 watts.

\subsection{Energy and Performance with the Generated Optimization Sequences}

We present next plots of energy/performance improvement ratios for both the dual Xeon and the ODROID platforms.
The ratios are calculated for the use of the generated compiler phase orders over the best compilation/execution configurations for energy, as depicted in Table \ref{table-bestoptimandthreads}.
For instance, for \emph{gemver} on the Xeon, the best configuration uses --O3 with 32 threads, so the results that we show next are ratios over that configuration for the use of the generated phase orders when compiling with OpenMP and executing with 32 threads.
Results falling over a straight line represent compiler phase orders that resulted in directly correlated energy consumption and performance (increases in energy efficiency lead to proportional increases in performance, and vice versa). 
Points diverging to one side or the other represent compiler sequences that lead to non-aligned energy efficiency or performance gains. 

\subsubsection{Dual Xeon}

\begin{figure*} [!t]
\setlength\abovecaptionskip{0pt}
 \begin{center}
  \includegraphics[width=7in]{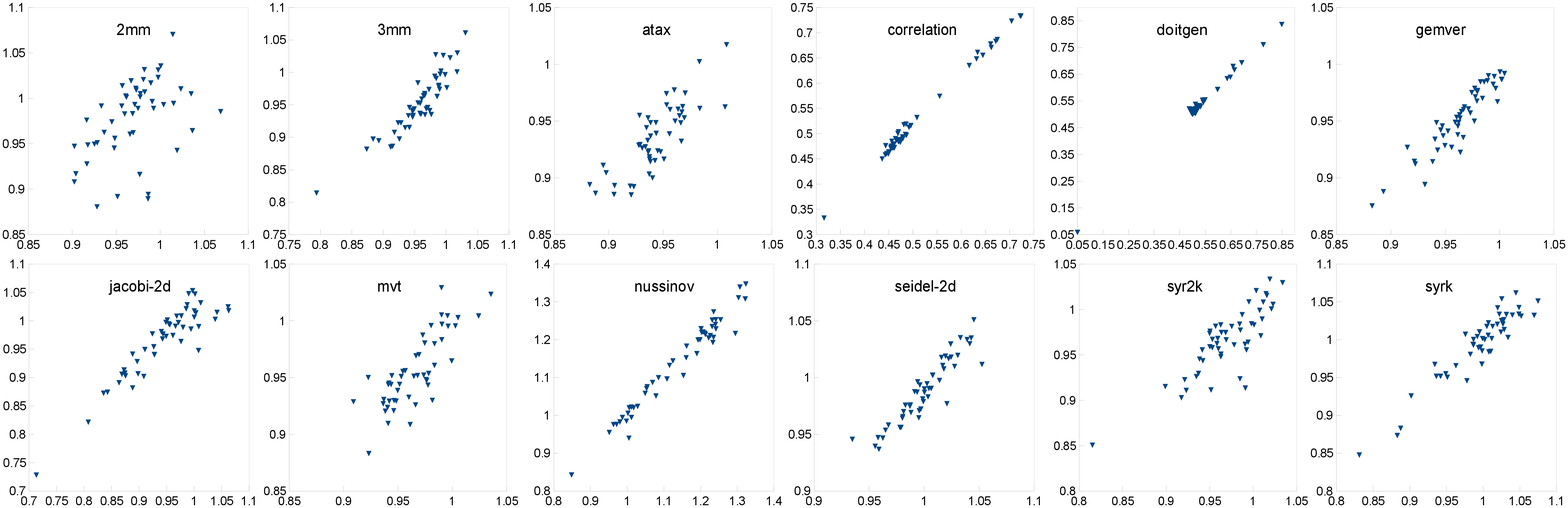}
 \end{center}
 \caption{Ratios of energy consumption (horizontal axis) and execution time (vertical axis) over the best per-function standard optimization level and execution configuration on the dual Xeon.}
 \label{fig-xeonbest}
\end{figure*}

Figure \ref{fig-xeonbest} shows energy consumption and execution time ratios for the binaries generated with phased orders specialized for improving energy efficiency.
The results presented are ratios over the best individually found --OX (i.e., --O0, --O1, --O2, or --O3) for each function, metric and target/configuration triplet.
The data collected from the experiments suggests that although energy efficiency and performance tend to correlate, there are sequences that lead to cases of performance gains without energy improvements, or vice versa.
For instance, in the case of \emph{2mm}, there are a lot of energy/performance ratios all over the chart, meaning there are a lot of sequences resulting in asymmetrical energy reduction and performance gains.
Additionally, for \emph{2mm} the compiler sequence that leads to best energy savings (point most to right) does not lead to the highest performance (point most to the top).
For \emph{atax} two compiler phase orders result in the same best energy efficiency (the two points most at right) while having different performance.
For the functions where the single threaded non-OpenMP version resulted in best energy efficiency, such as \emph{correlatio, doitgen}, the points on the chart form a straight line.

\subsubsection{ODROID XU+E}

\begin{figure*} [!t]
\setlength\abovecaptionskip{0pt}
 \begin{center}
  \includegraphics[width=7in]{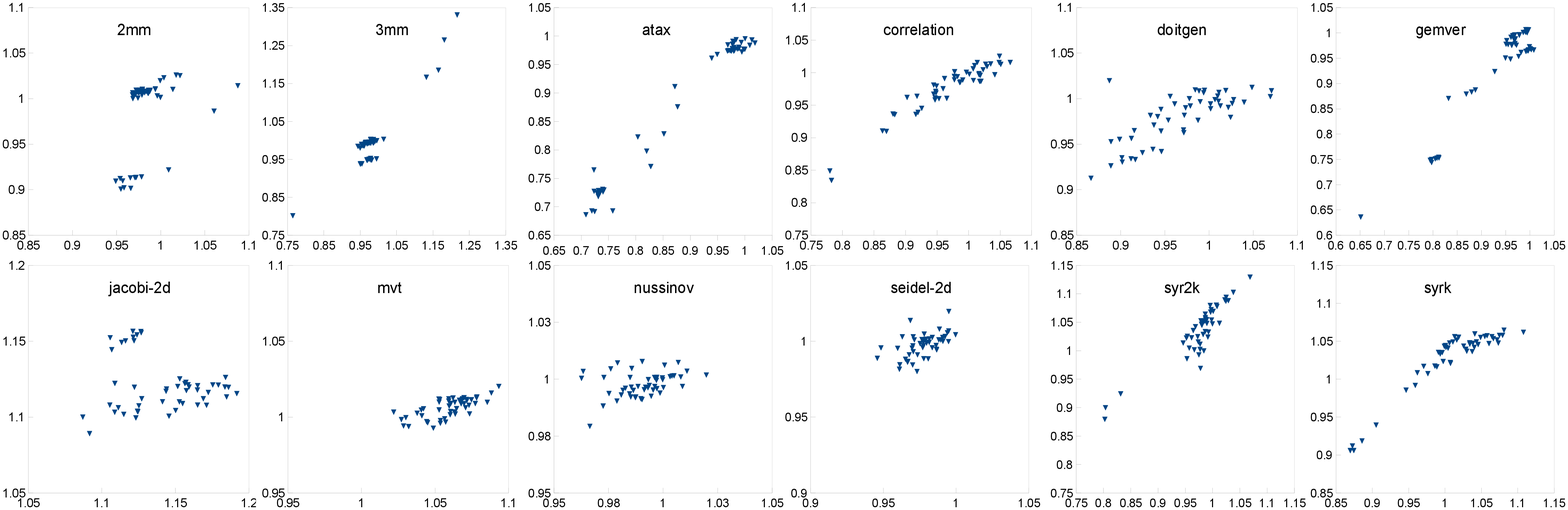}
 \end{center}
 \caption{Ratios of energy consumption (horizontal axis) and execution time (vertical axis) over the best per-function standard optimization level and execution configuration on the ODROID.}
 \label{fig-xeonbest}
\end{figure*}

As with the Xeon platform, although energy consumption and performance are for most part highly correlated (i.e., both tend to improve or get worse together), there are pairs of points (representing ratios of energy efficiency and performance using specialized phase orders) for any functions where energy decreases without performance increasing at all, and vice versa.
For instance, with \emph{2mm}, \emph{doitgen}, \emph{jacobi-2d}, \emph{nussinov}, \emph{mvt} and \emph{syrk} there are a number of sequences that result in considerably more energy reduction than impact on performance.

\section{Related Work} \label{section-relatedwork}

Manually devising suitable phase orderings requires deep knowledge about correlation between code features, target architecture and compiler passes interdependence.
Compilers typically have a large number of compiler passes, making this a complex problem.

Testing all possible passes combinations is not feasible in many cases (e.g., LLVM has more than a hundred of passes), as that would result in $P^K$ phase orders if considering $P$ passes and phase orders composed of up to $K$ passes each.
To be able to perform automatic phase ordering exploration despite the large design space, researchers have been proposing a number of approaches, falling in the iterative and/or ML-based category.
Most only have performance as optimization metric when considering software compilation, which may have to be in part related to the fact that most simulators and execution platforms do not provide an integrated way to measure energy and/or power.

Almagor et al. \cite{almagor2004} rely on Genetic Algorithms (GAs), hill climbers, and greedy algorithms to explore compiler phase ordering at program-level.
With 200 to 4,550 compilations, their approach can find custom sequences that are 15\% to 25\% better than the human-designed fixed sequence originally used by the compiler when targeting a SPARC processor.

Kulkarni et al. \cite{kulkarni2004} propose GAs to iteratively explore compiler pass sequences for improving performance at function-level, targeting an Intel StrongARM SA-100 processor.
In this work, 15 compiler passes 
of the Very Portable Optimizer were considered for exploration.
Two approaches for achieving faster searches when using GAs are presented.
They improve exploration efficiency by avoiding unnecessary executions and modified the search, resulting in average search time reductions of 62\% and in a reduction of average GA generations by 59\%.
Additional techniques to prune the exploration space are presented in \cite{kulkarni2009} and \cite{kulkarni2010}.

Copper et al. \cite{cooper2006} explore phase orders at program-level with randomized search algorithms based on genetic algorithms, hill climbers and randomized sampling.
They target a simulated abstract RISC-based processor with a research compiler, and report properties of several of the generated sub-spaces of phase ordering and the consequences of those properties for the search algorithms.

Agakov et al. \cite{agakov2006} present a methodology to reduce the number of evaluations of the program.
Models are generated taking into account program features (30 features reduced to 5 using principal component analysis) and the shapes of compiler sequence spaces generated from iteratively evaluating a training set of programs.
These models are then used by the iterative exploration for a new program.
They present results concerning the evaluation of two distinct models, an independent identically distributed model and a stationary Markov model, when compiling with the SUIF source-to-source compiler coupled with Code Composer and GCC, for generating code for the TI C6713 and AMD Au1500 embedded processors.
The two models were tested with GAs in order to determine how much of the design space can be pruned by the proposed approach.
Experimental results using the leave-one-out method show the exploration process can be accelerated by an order of magnitude, with no negative impact on the performance of the generated code.

Purini et al. \cite{purini2013} present an approach which relies on a list of compiler sequences previously found for a representative set of programs.
Given a new program, each of these compiler sequences is tested and the one leading to better performance is used to compile the new program.
The approach is tested considering 62 machine-independent LLVM 3.0 compiler passes when generating the list of compiler sequences considered for testing with new programs.
Results show an average speedup up to 14\% when targeting an Intel Xeon W35550.

Sher et al. \cite{sher2014} describe a compilation system that relies on evolutionary neural networks for phase ordering exploration using LLVM. 
The neural networks output a set of probabilities of use for each compiler pass, which is then sampled a number of times to generate different compiler sequences.
The neural networks use 48 and 44 features as input for the program- and the function-level approaches, respectively.
The system was able to find compiler sequences resulting in performance improvements between 5\% and 50\% on Intel Core i7 considering 53 (program-level) and 34 (function-level) LLVM compiler passes for exploration. 

Martins et al. \cite{martins2014, taco2016} proposed a clustering method to reduce the exploration space in the context of compiler pass phase selection and order exploration.
Performing clustering on top of source code representations generated with a fingerprinting method allows the classification of a new source code into one of the existing clusters.
Each cluster has associated with it only the compiler passes that are known to perform well with codes that are represented by similar fingerprints, so that the exploration space (and as direct result, the exploration time) is considerably reduced.
The approach explored the use of 49 compiler passes of the CoSy-based REFLECTC \cite{nobre2013book} compiler and of 124 passes when considering the use of LLVM 3.5 \cite{llvm}.
Experimental results reveal that the clustering-based DSE approach achieved a significant reduction on the total exploration time of the search space (18$\times$ over a Genetic Algorithm approach for DSE) at the same time that important performance speedups (43\% over the baseline) were obtained by the optimized codes.

In a previous work \cite{lctes2016} we presented and validated an approach that uses a graph to represent compiler pass orderings likely to improve performance of the generated binaries.
That graph is created based on statistical information extracted from sequences previously found for a reference set of functions.
In this paper we use a similar approach to generate compiler sequences focusing on reducing energy consumption.

\section{Conclusion}\label{section-conclusions}

This paper evaluated compiler pass phase ordering as an orthogonal approach to reduce energy consumption. The experiments presented in this paper include execution time, energy and power consumption measurements when considering different optimizations and OpenMP implementations (varying the number of threads) for both an ARM-based embedded system (ODROID) and an Xeon-based workstation. 

Our experimental results suggest that energy efficiency and performance are not perfectly correlated. 
Optimizing specifically for energy allows in some cases to achieve higher energy savings than if optimizing only for performance.
For the functions considered and for both the ODROID and the Xeon workstation, we found compiler phase orders that, in comparison with other phase orders, achieve higher energy savings while not improving (or even degrading) performance.

We are currently focusing on refining the process of building the graph model used to generate specialized phase orders and we are exploring approaches to multiobjective optimization.

The experiments presented in this paper can be extended by, e.g., testing the impact of different power levels on the Xeon workstation, repeating the experiments with the ODROID using the \emph{big.LITTLE} ARM CPU in different configurations (e.g., only the Cortex-A7 cores or only the Cortex-A15 cores).

\section*{Acknowledgments}

This work was partially supported by the TEC4Growth project, "NORTE-01-0145-FEDER-000020", financed by the North Portugal Regional Operational Programme (NORTE 2020), under the PORTUGAL 2020 Partnership Agreement, and through the European Regional Development Fund (ERDF), and by the ANTAREX FET-HPC H2020 Project (grant agreement No 671623).


\newpage{\pagestyle{empty}\cleardoublepage}

\end{document}